\input harvmac
\noblackbox
\overfullrule=0pt
\def\Title#1#2{\rightline{#1}\ifx\answ\bigans\nopagenumbers\pageno0\vskip1in
\else\pageno1\vskip.8in\fi \centerline{\titlefont #2}\vskip .5in}

scaled\magstep3
 
scaled\magstep3

\def\frac#1#2{{\textstyle #1 \over \textstyle #2}}
\def\apm{{\alpha^{\prime}}}

%
\lref\sone{N. Seiberg, hep-th/9606017.}
\lref\sw{N. Seiberg and E. Witten, hep-th/9607163.}
\lref\sfive{N. Seiberg and S. Shenker, 9608086.}
\lref\aks{M. Abe, H. Kubota and N. Sakai, {\sl Phys.~Lett.~B},
{\bf 200} (1988) 461.}
\lref\fesat{S.~Ferrara and S.~Sabharwal, {\sl Class.~Quantum Grav.}, 
{\bf 6} (1989) 77.}
\lref\fesa{S.~Ferrara and S.~Sabharwal, {\sl Nucl~Phys.~B},
{\bf 332}, (1990) 317.}
\lref\bcf{ M.~Bodner, A.C.~Cadavid, and S.~Ferrara, {\sl Class.~Quantum
Grav.}, {\bf 8} (1991) 789.}
\lref\ccaf{A.C.~Cadavid, A.~Ceresole, R.~D'Auria, and S.~Ferrara;
hep-th/9506144.}
\lref\bbsup{K.~Becker, M.~Becker, and A.~Strominger, unpublished.}
\lref\asup{A.~Strominger, unpublished.}
\lref\gsw{M.B.~Green, J.H.~Schwarz, and E.~Witten, {\sl Superstring
Theory II}, Cambridge Univ.~Press (1987).}
\lref\bbs{K.~Becker, M.~Becker, and A.~Strominger, {\sl Nucl~Phys.~B},
{\bf 456} (1995) 130.}
\lref\grgu{M.B.~Green and M.~Gutperle; hep-th/9701093.}
\lref\cgpo{P.~Candelas. X.C.~de la Ossa, P.S.~Green, and L.~Parkes, {\sl
Nucl.~Phys.~B}, {\bf 359} (1991) 21.}
\lref\hb{J.~Blum and J.~Harvey, unpublished.}
\lref\gw{D.J.~Gross and E.~Witten, {\sl Nucl~Phys.~B}, {\bf 277} (1986)
1.}
\lref\gv{M.B.~Green and P. Vanhove, hep-th/9704145.}
\lref\gs{M.B.~Green and J.H.~Schwarz, {\sl Nucl.~Phys.~B}, {\bf 198}
(1982) 441.}
\lref\st{N.~Sakai and Y.~Tanii, {\sl Nucl.~Phys.~B}, {\bf 287}, (1987)
457.}
\lref\rsw{M.~deRoo, H.~Suelmann, and A.Wiedemann, {\sl Phys.~Lett.~B},
{\bf 280} (1992) 39.}
\lref\tse{A.A.~Tseytlin, {\sl Nucl~Phys.~B}, {\bf 467} (1996) 383.}
\lref\duff{M.J.~Duff, J.T.~Liu, and R.Minasian, {\sl Nucl~Phys.~B}, {\bf
452} (1995) 261.}
\lref\vw{C.~Vafa and E.~Witten, {\sl Nucl.~Phys.~B}, {\bf 447} (1995)
261.}
\lref\cfg{C.~Cecotti, S.~Ferrara, and L.~Girardello, {\sl
Int.~J.~Mod.~Phys.~A}, {\bf 4}, (1989) 2475.}
\Title{\vbox{\baselineskip12pt
\hbox{UCSBTH-97-13}\hbox{hep-th/9706195}}}
{\vbox{\centerline{\bf{ LOOP CORRECTIONS TO THE}}
\vskip2pt\centerline{\bf{UNIVERSAL HYPERMULTIPLET}}}}

{
\baselineskip=12pt
\centerline{ Andrew Strominger }
\smallskip
\centerline{\sl Department of Physics}
\centerline{\sl University of California}
\centerline{\sl Santa Barbara, CA 93206-9530}
\bigskip
\centerline{\bf Abstract}
The universal hypermultiplet arises as a subsector of every 
Calabi-Yau compactification of $M$-theory or Type II string theory.
Classically its moduli space is the quaternionic space $SU(2,1)/U(2)$.
We show that this moduli space 
receives a one-loop correction proportional to 
the Euler character of the Calabi-Yau, which can locally 
be absorbed by a certain constant
shift of the fields. The correction vanishes in the limit
that the Planck mass is taken to infinity, and hence is essentially 
gravitational in nature.
}

\Date{}

\newsec{Introduction}

In the last several years there has been a great deal of progress in
understanding quantum corrections to moduli spaces of string vacua.
Most of the results can be understood in a field theory limit in which
gravity is decoupled by taking the string or Planck mass to infinity. However,
it is possible that there are qualitatively new quantum phenomena which
occur only when gravity is included.

The universal hypermultiplet of $N=2$ string compactifications \cfg \ is an
interesting place to search for such phenomena. This arises as a
subsector in every $N=2$ Calabi-Yau string compactification and classically
parameterizes the quaternionic space $SU(2, 1)/U(2)$. Under type II
mirror symmetry it transforms into the gravity multiplet rather than a
vector multiplet. When the
Planck mass is taken to infinity, the curvature of $SU(2, 1)/U(2)$ goes
to zero, and it reduces to a free supermultiplet. There are then no
quantum corrections.  Hence any corrections to the classical moduli
space must be essentially gravitational in nature\foot{Quantum 
corrections to 
hypermulitplet moduli spaces have been analyzed 
in essentially field-theoretic contexts in \refs{\sone \sw - \sfive}.}.

In this paper we shall show, in the context of $M$-theory or IIA 
Calabi-Yau
compactification, that the moduli space of the universal hypermultiplet
indeed gets a one-loop correction proportional to the Euler character
$\chi$ of the Calabi-Yau space. These are derived from $(Riemann)^4$ 
terms which have been the subject of interesting recent investigations
\grgu, \gv. We shall further find an exactly 
quaternionic metric which reproduces the tree-level and 
one loop results and has 
non-trivial corrections at every order in perturbation theory. 
This exact metric is in fact related to the original metric by 
a field redefinition of the 
form $1/g^2 \to constant\cdot \chi +1/g^2$. Hence, 
locally the geometry remains 
$SU(2,1)/U(2)$. However globally string duality implies 
a number of discrete identifications \bbsup. 
When these are taken into account,
the shift may have nontrivial consequences.

Looking beyond the scope of this paper, in \bbs\ it was 
argued that the hypermultiplet 
metrics in general are $\it non-perturbatively$ corrected 
by both membrane and fivebrane instantons\foot{In 
\grgu, \gv,  nonperturbative D-instanton corrections in 
ten-dimensional IIB theory which should be relevant to this issue were  
found. }. A complete nonperturbative picture of the 
quantum moduli space for the universal hypermultiplet 
potentially includes such corrections as well as  the perturbative 
shift discussed herein and global identifications.

In section 2 we describe how the universal hypermultiplet arises 
for $M$-theory or IIA compactification on a rigid Calabi-Yau. In
section 3 one-loop $(Riemann)^4$ corrections in $M$-theory are 
discussed. In section 4 it is shown that upon compactification 
this one-loop term corrects one component of the metric on the 
hypermultiplet moduli space. In section 5 the quaternionic 
geometry of
the universal hypermultiplet is reviewed. In section 6 the results
of section 4, together with the constraints 
of quaternionic geometry and some assumptions about the symmetries of 
string perturbation theory, are used to deduce the complete form of the
one-loop correction (with some details relegated to an appendix). 
The quaternionic constraints imply that 
the corrections can not terminate at one loop. In section 7 we present
the fully corrected metric and explain how the correction can be locally
absorbed by a field redefinition.   

\newsec{The Universal Hypermultiplet in IIA String Theory and $M$-Theory}

Compactification of $M$-Theory or  IIA string theory on a rigid
Calabi-Yau ($h_{21}=0$) leads to an $N=2$ theory with a single universal
hypermultiplet \cfg, \refs{\fesat \fesa \bcf \ccaf - \bbsup} parameterized by
the complex fields $S$ and $C$.
The leading low-energy action for $M$-theory contains the terms
\eqn\twoone{S^0_M = \frac{1}{2} \int\, d^{11} x \sqrt{-g}
\ R -\frac{1}{4}\ \int\, [F\wedge * F + \frac{1}{3} A_3 \wedge
F\wedge F],}
where $F=dA_3$.
In the $M$-theory case, the compactified 
theory is five-dimensional. The real part of
$S$ is related to the volume $V_M$ of the Calabi-Yau by 
\eqn\slvn{ReS = {V_M}\equiv e^{2D}.}
We will be particularly interested in the kinetic term for $D$.
After rescaling the five-dimensional metric by a factor of $e^{4D/3}$
to the Einstein frame, 
one finds a term
\eqn\whtt{-\int d^5x\sqrt{-g}(\nabla D)^2.}
The imaginary
part of $S$ derives from the dual of the four-form $F=dA_3$, and $C$
corresponds to expectation values for $A_3$ proportional to the
holomorphic three-form of the Calabi-Yau. The full $S, C$ metric will 
be given in section 5 below. 

For IIA compactification on a rigid Calabi-Yau, one classically obtains
the same universal hypermultiplet in four dimensions.
Now, however, one finds
\eqn\stcn{ReS=e^{-2\phi_4},}
where $\phi_4$ is the four-dimensional string dilaton.  It is related to
the ten-dimensional string dilaton $\phi_{10}$ and the
string frame volume $V_{str}$ by
\eqn\twofour{e^{-2\phi_4} = e^{-2\phi_{10}} {V_{str}}\ .}
A second scalar field, the volume $V_{str}$, is part of a vector
multiplet.  The imaginary part of $S$ is related to the dual of the
three-form $H$, and $C$ is again proportional to $A_3$ expectation
values. $S$ is a $NS-NS$ field while $C$ is a $R-R$ field.

The $M$-theory metric $ds^2_M$ is related to the string metric
$ds^2_{str}$ by 
\eqn\mstr{ds^2_M = e^{4\phi_{10}/3} (dx^{11})^2 + e^{-2\phi_{10}/3} 
ds^2_{str},}
where $x^{11} \sim x^{11}+1$ and 
the IIA gauge field is suppressed. This implies
that the $S$ defined in $M$-theory \slvn\ and the $S$ defined in
the IIA theory \stcn\ are the same or, equivalently,
\eqn\twosix{\phi_4=-D\ .}
The five-dimensional $M$-theory
vacuum can be reached from the four-dimensional IIA vacuum by taking the
radius of the eleventh dimension, or $\phi_{10}$, to infinity while 
keeping the Calabi-Yau volume in the $M$-theory frame, or
$\phi_4$, fixed. This implies that $V_{str}$ must be taken to infinity.
Since $V_{str}$ is part of a vector multiplet, the $M$-theory limit of
IIA Calabi-Yau compactification is a boundary in the vector multiplet
moduli space.  Since neutral hyper and vector multiplets decouple, this
implies that the same hypermultiplet moduli space is obtained for either
IIA or $M$-theory on a given Calabi-Yau. Furthermore, the four-dimensional IIA
dilaton $\phi_4$ becomes the $M$-theory six-volume $D$.  Hence the IIA
loop expansion will correspond to an expansion in higher dimension
operators in the eleven-dimensional $M$-theory action.

The $M$-theory approach is in some ways simpler because the problem of
untangling the radial and string dilaton is avoided. On the other hand,
corrections are more readily calculated in the IIA picture. We shall
find it useful to use both pictures in the following.

\newsec{$R^4$ Corrections in Ten and Eleven Dimensions}

The leading correction to the purely gravitational part of
the IIA action has been inferred from four graviton
scattering in \gw, \gs, \st. The corrected action 
is given by, in the string frame ($\apm=1$)
\eqn\threeone{{1\over 2 } \int\ d^{10}x \sqrt{-g}
\bigl[ e^{-2\phi_{10}}R -{ c_0\over 3\cdot 2^{6}}
\bigl( e^{-2\phi_{10}} 
\zeta (3) +c_1\bigr)Y \bigr],}
where we have included the tree-level $\zeta (3) $ term \gw\ for comparison. 
$Y$ here is the quartic curvature invariant defined by
\eqn\threetwo{\eqalign{Y &\equiv \hat t_8 
\hat t_8 R^4 - \frac{1}{4} \varepsilon_{10}
 \varepsilon_{10} R^4,\cr
\hat t_8 \hat t_8 R^4 &\equiv \hat t^{\mu_1\nu_1\cdots\mu_4\nu_4}
\ \hat t_{\alpha_1\beta_1\cdots\alpha_4\beta_4}
 R_{\mu_1\ \nu_1}^{ \ \alpha_1\ \beta_1} 
\cdots R_{\mu_4\ \nu_4}^{\ \alpha_4\ \beta_4},\cr
\varepsilon_{10} \varepsilon_{10}\ R^4 &\equiv
\varepsilon^{\rho\sigma\mu_1\nu_1\cdots \mu_4\nu_4}
\ \varepsilon_{\rho\sigma\alpha_1\beta_1 \cdots \alpha_4\beta_4}
R_{\mu_1\ \nu_1}^{\ \alpha_1\ \beta_1} 
\cdots R_{\mu_4\ \nu_4}^{\ \alpha_4\ \beta_4}.\cr}}
The tensor $\hat t$ can be found in Appendix 9A of \gsw. 
According to \gv\ the constants 
$c_0$ and $c_1$ are given by\foot{In \gv\ $\apm =1$ 
and the dilaton is shifted so that
$e^{-2\phi_{GV}}=e^{-2\phi}2^6\pi^7$.} 
\eqn\czro{c_0=1,}
\eqn\cone{c_1={1\over 3\cdot 2^6 \pi ^5}.}
There appears to be unresolved discrepancies in 
the values of these constants in the literature 
(see \st, \aks, \tse, \grgu, \gv\ for discussion). For now we will
simply quote our results in terms of $c_0$ and $c_1$.

$Y$ also arises at one loop for the heterotic string \st. In that
context it was shown \rsw\ to be part of an $N=1$ supermultiplet of terms
containing the anomaly-canceling term $\int B\wedge trR\wedge R\wedge
R\wedge R$. It has been plausibly argued \tse\ that in the type II
context it is also part of an $N=2$ supermultiplet of terms containing
$\int B\wedge tr R\wedge R\wedge R\wedge R$.

In the $M$-theory limit, $\int B\wedge tr R\wedge R\wedge R\wedge R$
goes over to $\int A_3 \wedge tr  R\wedge R\wedge R\wedge R$. Because
this term is connected to an inflow anomaly, the coefficient does not
change in the transition from IIA to $M$-theory \hb, \duff, \vw, \gv.
In terms of the $M$-theory metric defined in \mstr, the 
corrected action becomes 
\eqn\thrfive{{1\over 2 } \int\ d^{10}x \sqrt{-g}
\bigl[ e^{2\phi_{10}/3}R -{ c_0\over 3\cdot 2^{6}}
\bigl( e^{-4\phi_{10}/3} 
\zeta (3) +{e^{2\phi_{10}/3} c_1 } \bigr)Y \bigr],}
where we have suppressed terms involving derivatives of $\phi_{10}$.
The prefactor $e^{2\phi_{10}/3}$ is just the radius of the eleventh
dimension,  so the $M$-theory action contains the terms (as 
similarly observed in \gv )
\eqn\threefi{{1\over 2 } \int\ d^{11}x \sqrt{-g}
\bigl[ R -{ c_0c_1 \over 3\cdot   2^{6} } Y \bigr].}
In general one cannot naively extrapolate coefficients from IIA to
$M$-theory in this fashion, but in this case the coefficient is
protected by the anomaly.
Note that the $\zeta (3) $ term 
vanishes in the $M$-theory limit.

\newsec{Compactification}

In this section we consider the compactification of 
the $M$-theory terms \threefi\ on a Calabi-Yau space.
In the context of tree-level string compactification 
it has been shown \cgpo\ that the quartic invariant $Y$ 
leads to a correction to the kinetic term for the scalar
$D$ governing the size of the Calabi-Yau.  To see how this 
arises 
consider the ansatz
\eqn\fourone{d s^2 = e^{2D(x^a)/3} \bar g_{MN} dx^M dx^N + \eta_{ab}
 dx^a dx^b,}
where $a, b= 0, \cdots, 4\ ; M,N = 5, \cdots, 10\ ,$
$\bar g_{MN}$ is the Ricci flat metric on the unit-volume 
Calabi-Yau and $D$ is
allowed to depend on the Minkowski space coordinate $x^a$. The Riemann
tensor then becomes
\eqn\rtns{\eqalign{R_{\mu\nu\lambda}\ ^\rho & = \bar R_{\mu\nu\lambda}
\ ^\rho - 2 \nabla_{[\mu} C^\rho_{\nu]\lambda} + 2 C^\sigma_{\lambda[\mu}
C^\rho_{\nu]\sigma},\cr
C^\mu_{\nu\lambda} & \equiv {1\over 3}\bigl( \delta^\mu_\nu \nabla_\lambda D +
\delta^\mu_\lambda \nabla_\nu D- g_{\nu\lambda} g^{\mu\rho} 
 \nabla_\rho D\bigr),\cr}}
where $\mu, \nu=0,\cdots, 10$ and $\bar R$ is the Rienman tensor for
$D=0$.
We are interested in terms descending from $Y$ proportional to
$(\nabla D)^2$. These will involve the last term in the expansion \rtns\ of
the Riemann tensor and three powers of ${{{\bar R}_{MNP}}}\ ^Q$. 
Hence $Y$ will yield terms of the form 
\eqn\fourfour{ \nabla^a D\nabla_a D \bar R_{MN}
\ ^{PQ}\bar R_{PQ}\ ^{RS}\bar R_{RS}\ ^{MN}\ .}
Integrating over the Calabi-Yau and rescaling to the Einstein 
frame  yields the loop-corrected action 
\eqn\dltl{ -\int d^5x\sqrt{-g}
\bigl(\nabla D)^2(1-{c_1 \chi 
\over 40    \pi^3}e^{-2D}\bigr),}
where $\chi$ is  the Euler character of the Calabi-Yau.

In order to determine the correction in
\dltl\ we have employed a shortcut rather 
than the direct procedure outlined above. In \cgpo\ it was shown 
in the context of string theory compactification 
that the tree-level term corrects the metric for $D$ by a factor 
of\foot{In the conventions of \cgpo\ $6e^{2D}=t^3$.} 
\eqn\cpcr{1-{ \zeta (3) \chi \over 40 \pi^3}e^{-2D}.}
Comparing the coefficients of the one-loop and tree-level terms 
in \threeone\ then yields \dltl.

\dltl\ is of course not the only one-loop correction to the metric.
Corrections to other components could arise for example from terms of
the form $F^2R^3$ in eleven dimensions.  It appears difficult to
determine their coefficients directly.  However, in section 6 we shall
fix the full one-loop result from symmetries and constraints of
quaternionic geometry. A direct IIA calculation, although possible in
principle, is tedious because one must compute terms like $(\nabla
\phi_{10})^2 R^3$. These are determined from one-loop five point
functions and are required in order to untangle the radial dilaton from the
universal hypermultiplet.

\newsec{Review of $SU(2,1)/U(2)$ and Quaternionic Geometry}

An $n$-dimensional quaternionic 
space has an $Sp(1)$ triplet of almost complex structures obeying
\eqn\fiveone{
J^i_a\ ^b\ J^j_b\ ^c  = - \delta^{ij}\delta_a\ ^c +
\varepsilon^{ij k}\ J^k_a\ ^c,}
where $i,j = 1,2,3$ and $a,b=1,\cdots, 4n$. $J^i$ together with
the metric define a triplet of two forms
\eqn\fivetwo{\Omega^i = \half J^i_a\ ^b\ g_{bc}\ dx^a\wedge dx^c\ .}
The holonomy group arising from $g$ is $Sp(1) \otimes Sp(n)$.
$\Omega^i$ is the curvature of the $Sp(1)$ connection $p$
\eqn\ccca{dp^i + \half \varepsilon^{ij k}\ p^j\wedge p^k = \Lambda\Omega^i\ .}
In our conventions $\Lambda = 1$. In contrast, for a
hyperkahler geometry the $Sp(1)$ curvature vanishes which corresponds
to
$\Lambda = 0$. In a more general set of conventions $\Lambda$ is
Newton's constant, and one sees directly that the quaternionic structure
becomes hyperkahler when gravity is turned off.  

For the universal hypermultiplet $n=1$ and the classical quaternionic space
is (locally) $SU(2,1)/U(2)$. The metric is explicitly
\eqn\fivefour{ds^2= \bar u u + \bar v v,}
where
\eqn\fivefive{\eqalign{
u & \equiv e^\phi\ dC\ ,\cr
v & \equiv e^{2\phi}\ \left(\frac{dS}{2} - \bar C\ dC\right)\ ,\cr
\phi & \equiv - \half\ \ell n\left[(S+\bar S - 2 C \bar C)/2\right]
\ .\cr}}
$\phi$ here is $\phi_4$ or $-D$ with $R-R$ corrections. 
It is convenient to introduce the vierbein
\eqn\fivesix{V=\pmatrix{~~u\cr
           ~~v\cr
           ~~\bar v\cr
           -\bar u\cr}}
with metric 
\eqn\fiveseven{
\half \sigma_2 \otimes \sigma_2 = \pmatrix{0&0&0&-\half\cr
                                       0&0&\half&0\cr
                                       0&\half&0&0\cr
                                       -\half&0&0&0\cr}.}
The holonomy group $O(4)$ decomposes as $Sp(1) \otimes Sp(1)^\prime$
with connections $p$ and $q$. $p$ acts on $v$ as
\eqn\fiveeight{p= -{i \over 2}p^i\pmatrix{\sigma^i&0\cr
                                 0&\sigma^i\cr} = -{i \over 2}
p^i\sigma^i \otimes
{\bf 1}_2 ,}
so that $u$ and $v$ are an $Sp(1)$ doublet. The second $Sp(1)^\prime$
connection $q$ acts as ${\bf 1}_2 \otimes q^k\sigma^k$, 
and commutes with $p$. We are primarily interested in $p$,
rather than $q$, because it is subject to the constant curvature
constraint \ccca. $p$ and $q$ are determined from $V$ by
\eqn\pqv{dV + p\wedge V + q\wedge V=0\ .}
One finds
\eqn\fiveten{p+q=\pmatrix{\half (\bar v-v)&-u&0&0\cr
                              \bar u&\bar v-v&0&0\cr
                              0&0&v-\bar v&-u\cr
                              0&0&\bar u&\half(v-\bar v)\cr}.}
The two-form triplet $\Omega^i$ is given by
\eqn\wvd{\Omega^i = {i \over 2} \bar V \wedge \Sigma^i V,}
where $\Sigma^i \equiv \sigma^i \otimes {\bf 1}_2 $ and 
$\bar V$ is constructed with \fiveseven. Explicitly
\eqn\fivetwelve{\eqalign{
\Omega^1 & = {i }(\bar u\wedge v + \bar v \wedge u),\cr
\Omega^2 & = (\bar u\wedge v - \bar v \wedge u),\cr
\Omega^3 & = {i }(\bar u\wedge u - \bar v \wedge v).\cr}}
One may verify that
\eqn\ccca{dp^i+\half \varepsilon^{ijk}p^j \wedge p^k = \Omega^i,}
and the geometry is therefore quaternionic.

Some useful relations are
\eqn\fivethirteen{\eqalign{
dv & = v\wedge \bar v + u \wedge \bar u,\cr
du & = \half u \wedge (v+\bar v),\cr
d\phi & = -\half (v + \bar v).\cr}}

\newsec{Quaternionic Perturbations}

The $SU(2,1)/U(2)$ vierbein  in \fivesix\ cannot be the exact answer
because at one loop the correction \dltl\ is encountered.  In addition
to \dltl\ there may be further one-loop corrections to other metric
components.  These should combine into a linearized quaternionic
perturbation. A perturbation 
$\delta V$ is quaternionic to first order if and only if
the variation $\delta \Omega^i$ of $\Omega^i$ induced from \wvd\ and the
variation $\delta p^i$ of $p^i$ induced from \pqv\ are related by the
linearization of \ccca:
\eqn\qcd{\eqalign{
d\delta p^i + \half \varepsilon^{ij k}\ \delta p^j \wedge p^k
            + \half \varepsilon^{ij k}\ p^j \wedge \ \delta
p^k = \delta \Omega^i\ .}}
$\delta V$ is further constrained by the following observations:
\item{(1)} At string tree level there are three relevant
Peccei-Quinn symmetries corresponding to constant shifts of the $NS-NS$
axion and the two $R-R$ three-form potentials. These act
as\foot{Charge quantization implies discrete identifications along
these shifts \bbsup.}
\eqn\pqs{\eqalign{
S & \to S+ i\theta + 2 \bar \varepsilon C,\cr
C & \to C + \varepsilon,\cr}}
where $\theta$ is real and $\varepsilon$ is complex, and generate 
a subgroup of $SU(2,1)$. Note that $u, v$
and $\phi$ are all invariant under \pqs. 
We have not proven but will assume that these symmetries 
are unmodified in string perturbation theory and hence 
$\delta V$ must be invariant as
well.
\item{(2)} Perturbative string amplitudes with an odd number of $R-R$
fields vanish.  Hence all terms in the perturbed metric must have an
even number of $R-R$ fields. This rules out for example $\bar u v$ 
corrections. 
\item{(3)} The perturbative theory is parity invariant. 
Parity changes the sign of the $NS-NS$ axion, and hence rules out 
$vv$ corrections to the metric. 
\item{(4)} As we are interested in one-loop corrections, $\delta V$
should scale like $\lambda^{-2}$ under $S\to\lambda^2S, C\to\lambda C$.
\item{(5)} The coefficient of $(\nabla D)^2$ must agree with \dltl.

One may verify the following perturbations obey the linearized 
quaternionic relations and are consistent with (1)-(4)
\eqn\apne{\delta V= e^{2\phi}\pmatrix{u\cr
                                       2v\cr
                                       2\bar v\cr
                                       - \bar u\cr}}
\eqn\sixfive{\delta p = e^{2\phi}\pmatrix{\half(v-\bar v)&-u&0&0\cr
                                  \bar u& \half(\bar v-v)&0&0\cr
                                  0&0&\half(v-\bar v)&-u\cr
                                  0&0&\bar u&\half( v-\bar v)\cr},}
\eqn\sixsix{\delta q =  e^{2\phi}\pmatrix{\frac{3}{2}(\bar
v-v)&0&0&0\cr
                                      0&\frac{3}{2}(\bar v-v)&0&0\cr
                                      0&0&\frac{3}{2}(v-\bar v)&0\cr
                                      0&0&0&\frac{3}{2}(v-\bar v)\cr},}
\eqn\sixseven{\eqalign{
\delta\Omega^1 & = {3i}{} e^{2\phi}(\bar u \wedge v + \bar v
\wedge u),\cr
\delta\Omega^2 & = {3} e^{2\phi} (\bar u \wedge v - \bar v
\wedge u),\cr
\delta\Omega^3 & = 2i(\bar u \wedge u - 2 \bar v \wedge v).\cr }} 
In the appendix we show that this is the unique perturbation 
consistent with (1)-(4). 
Finally matching to the computed coefficient of $(\nabla D)^2$ one finds
\eqn\sixeight{\delta V = -\frac{c_1\chi e^{2\phi}}{160  \pi^3}
                                                     \pmatrix{u\cr
                                                      2v\cr
                                                      2\bar v\cr
                                                      -\bar u\cr}\ .}
The quaternionic metric is then given, through one loop order, by 
\eqn\sixnine{ds^2 = \bar u u + \bar v v - \frac{c_1 \chi e^{2\phi}}{80 
\pi^3}\ (\bar u u + 2 \bar v v) + {\cal O}(e^{4\phi})\ .}
\newsec{All Orders in Perturbation Theory}
In this section we present an all-orders corrected metric 
and relate it to $SU(2,1)/U(2)$ by a field redefinition. Let 
\eqn\fiveur{d  s^{\prime 2}=  \bar u^\prime  u^\prime +  
\bar v^\prime  v^\prime,}
where
\eqn\fivefiv{\eqalign{
 u^\prime & \equiv {1 \over \sqrt{e^{-2\phi}+A}} dC\ ,\cr
 v^\prime & \equiv {1\over {e^{-2\phi}+A}} 
\left(\frac{dS}{2} - \bar C\ dC\right)\  ,\cr}}
and $A$ is a constant. 
This agrees with \fivefive\ when $A=0$. Comparison with \sixnine\ yields
\eqn\avl{A={c_1\chi \over 80 \pi^3}.}

\fivefiv\ amounts to a shift 
in the inverse coupling $e^{-2\phi}\to e^{-2\phi}+A$, or eqivalently
$S\to S+A$. Since this is just a coordinate transformation \fivefiv\
is obvioiusly still quaternionic and locally $SU(2,1)/U(2)$. 
When global identifications \bbsup\ are taken into account the 
effects of the shift may not be trivial.  
We leave this issue 
for future exploration.

\bigskip
\noindent {\bf Acknowledgments}

This work was supported in part by DOE grant DOE91ER40618.  I am
grateful to Katrin and Melanie Becker for collaboration at earlier stages
of this work. I would also like to thank Jeff Harvey, Jeremy Michelson,
Greg Moore, Joe Polchinski and Nati Seiberg for useful conversations.

\vfill

\appendix{A}{Linearized Quaternionic Perturbations}

In this appendix we show that
\eqn\apone{\delta V= e^{2\phi}\pmatrix{u\cr
                                       2v\cr
                                       2\bar v\cr
                                       - \bar u\cr}}
is the unique one-loop quaternionic perturbation, up to local $Sp(1)\otimes
Sp(1)^\prime$ rotations (which give phase transformations of $u$ and
$v$), consistent with the perturbative symmetries 
discussed in Section 5. 

Considerations (1), (2), (3) and (4) limit $\delta V$ to the general
form,
up to $Sp(1) \otimes Sp(1)^\prime$ transformations,
\eqn\vptr{\delta V= e^{2\phi}\pmatrix{au & + & \beta\bar u\cr
                                      av & \cr
                                     a\bar v & \cr
                                    -a\bar u & - & \bar \beta u\cr},}
where $a$ is real and $\beta$ is complex. One then 
finds\foot{The $\wedge$ symbol is suppressed in the following.}
\eqn\apthree{\eqalign{e^{-2\phi}\delta\Omega^3 & = 2i a (\bar u u - \bar
v v)\ ,\cr
                      e^{-2\phi}\delta\Omega^+ & = 2 i a \bar u v +
{i\bar\beta} uv ,\cr}}
where $\Omega^\pm = \half(\Omega^1 \pm i \Omega^2)$. The 
deformation of the $Sp(1)$
connection $\delta p$ is constrained by the linearized equations 
\eqn\apfour{\eqalign{d\delta p^+& + i\delta p^3 p^+ + i
 p^3 \delta p^+ = \delta\Omega^+\ ,\cr
d\delta p^3 & + 2 i\delta p^+ p^- + 2 i p^+\delta p^-
= \delta\Omega^3\ .\cr}}
Substituting the known quantities yields
\eqn\ope{d\delta p^+ - \delta p^3\bar u - \half (v-\bar v)
\delta p^+ = e^{2\phi} ( 2ia\bar u v +
i\bar \beta u v) \ ,}
and
\eqn\otre{d\delta p^3 + 2\delta p^+ u - 2 \bar u
\delta p^- = 
2iae^{2\phi}
(\bar u u - \bar v v) \ .}

First we show $\beta = 0$. The only way to obtain a $\bar\beta uv$
term on the RHS of \ope\ is if $\delta p^+$ has a term proportional
to $ue^{2\phi}$.  But then an unwanted $u\bar v$ term will appear
on the LHS, which is not on the RHS. Hence we conclude
\eqn\apseven{\beta = 0.}

Once $\beta=0$ \vptr\ reduces to a scale transformation which is 
clearly not quaternionic (for fixed $\Lambda$). To see this 
explicitly consider \otre. In order to reproduce 
the $\bar v v$ term on the LHS,
$\delta p_3$ must be of the form
\eqn\apeight{e^{-2\phi}\delta p_3 = \frac{ia}{2} (v-\bar v) + c_0 (v
+ \bar v) + c_1 u + \bar c_1 \bar u\ .}
where $c_0$ is real but $c_1$ is complex.  Cancellation of $\bar u u$
terms in \otre\ then requires
\eqn\apnine{e^{-2\phi}\delta p^+ = \frac{3ia}{4}\ \bar u + d_1 v + d_2
\bar v\ ,}
for $d_1, d_2$ complex.  Now let us consider $\bar u v$ 
and $\bar u \bar v$ terms in \ope.
They must obey 
\eqn\apten{-\frac{9ia}{8} (v + \bar v) \bar u - 
\frac{ia}{2} (v-\bar v) \bar u -
c_0 (v + \bar v) \bar u - \frac{3ia}{8} (v-\bar v) \bar u = 2 ia\bar u v.}
Cancellation of $\bar v\bar u$ terms implies $a=c_0=0$. Hence no perturbation of the form
\vptr\ can be quaternionic, and we conclude that the unique quaternionic
perturbation is given by \sixeight\ up to local $Sp(1) \otimes
Sp(1)^\prime$ transformations.

\listrefs
\bye